\documentclass[aps,prl,twocolumn,english,preprintnumbers,amsmath,amssymb,nofootinbib,superscriptaddress]{revtex4-1}

\pdfoutput=1

\usepackage[usenames,dvipsnames]{color}

\usepackage[latin1]{inputenc}
\usepackage{graphicx}
\usepackage{color}
\usepackage{epstopdf}

\usepackage{ulem}

\usepackage{bm}
\usepackage{bbm}
\usepackage{amsmath}
\usepackage{slashed}
\usepackage{esint}
\usepackage{amsfonts}
\usepackage{amssymb}
\usepackage{tabularx}
\usepackage{slashed}

\usepackage{dsfont}

\newcommand{\be}{\begin{eqnarray}}
\newcommand{\ee}{\end{eqnarray}}

\newcommand{\Tc}{T_{\text{c}}}
\newcommand{\pat}{\partial_t}

\usepackage{multirow}

\usepackage{babel}
\makeatother

\begin{document}

\title{{Delayed Magnetic Catalysis}}
\author{Jens Braun} 
\affiliation{Institut f\"ur Kernphysik (Theoriezentrum), Technische Universit\"at Darmstadt, 
D-64289 Darmstadt, Germany}
\affiliation{ExtreMe Matter Institute EMMI, GSI, Planckstra{\ss}e 1, D-64291 Darmstadt, Germany}
\author{Walid Ahmed Mian}
\affiliation{Institut f\"ur Kernphysik (Theoriezentrum), Technische Universit\"at Darmstadt, 
D-64289 Darmstadt, Germany}
\author{Stefan Rechenberger}
\affiliation{Institut f\"ur Kernphysik (Theoriezentrum), Technische Universit\"at Darmstadt, 
D-64289 Darmstadt, Germany}

\begin{abstract}

We study the effect of an external magnetic field on the chiral phase transition
in the theory of the {strong interaction by} means of a 
renormalization-group (RG) fixed-point analysis, relying
on only one physical input parameter, the strong coupling at a given large momentum scale.
To be specific, we consider the interplay of the RG flow of four-quark interactions
and the running gauge coupling. Depending on the temperature and
the strength of the magnetic field, the gauge coupling can drive the quark sector
to criticality, resulting in chiral symmetry breaking. In accordance with
lattice {Monte-Carlo} simulations, we find that the chiral phase transition temperature
decreases for small values of the external magnetic field. For large magnetic
field strengths, however, our fixed-point study predicts that the phase 
transition temperature increases monotonically.

\end{abstract}

\pacs{64.60.Ak, 11.15.-q}

\maketitle

{\it Introduction.--} {The dynamics of gauge theories is expected to be strongly affected 
by external magnetic fields.}
This is of great phenomenological relevance for a large variety of systems, ranging from 
condensed-matter theory~\cite{Gusynin:1994va,Gusynin}
over off-central heavy-ion collision experiments~\cite{Kharzeev:2007jp,Skokov:2009qp} to neutron stars~\cite{Duncan} 
and cosmological models~\cite{Vachaspati}. 
In fact, studies of the influence of an external magnetic field on the phase 
diagram of the theory of the strong interaction {(Quantum Chromodynamics, QCD)} and its equation of state have attracted a lot of 
attention in recent years. In particular, the peculiar {inverse}-catalysis effect is here of great 
interest~\cite{Bali:2011qj,*Bali:2012zg}. It is related to the observation that the chiral critical temperature decreases with increasing 
strength of the magnetic field, in contradistinction to purely fermionic models 
also known from condensed-matter theory where magnetic catalysis is observed, i.e. the critical temperature increases with increasing 
magnetic field strength~\cite{Gusynin:1994va,Gusynin:1994xp,*Gusynin:1995nb}. 

The observation of inverse catalysis in lattice Monte-Carlo (MC) studies of the chiral phase transition in QCD 
has come unexpected~\cite{Bali:2011qj,Bali:2012zg}. This is related to the fact that 
effective low-energy QCD models are commonly believed to describe correctly many features of 
the QCD phase diagram, at least on a qualitative level. Since these models are predominantly
purely fermionic models being close relatives to the Nambu--Jona-Lasinio model, which has
been originally constructed based {on analogies} to condensed-matter
theory~\cite{Nambu:1961tp,*Nambu:1961fr}, it also appeared natural to expect that only
magnetic catalysis is at {work in QCD~\cite{Osipov:2007je,Fraga:2008qn}, see Ref.~\cite{Shovkovy:2012zn,*Andersen:2014xxa} for reviews.}

{Various extensions of low-energy QCD models have been studied, 
ranging from the inclusion of the order-parameter for deconfinement as obtained from lattice MC
{simulations~\cite{Mizher:2010zb,*Gatto:2010qs,*Gatto:2010pt,*Kashiwa:2011js}}
to extensions beyond the mean-field 
{approximation~\cite{Fukushima,Skokov:2011ib,*Kamikado:2013pya,*Kamikado:2014bua}.}}
{Moreover, the effect of a magnetic field on the chiral
dynamics has been recently studied using Dyson-Schwinger equations (DSE)~\cite{Mueller:2014tea}.}
In any case, the very observation of magnetic catalysis is {found to be generic, even} if effects associated
with the chiral anomaly are taken into account~\cite{Osipov:2007je}.
On the other hand, it has been found that the parameters of {low-energy} models 
can be tuned such that inverse catalysis occurs at weak magnetic fields~\cite{Fraga:2013ova,Ferreira:2014kpa}.
Interestingly, {at strong magnetic fields,} catalysis is still observed and appears to be
a robust feature of these models.

The detailed analysis of low-energy models suggests that 
a formulation of the problem in terms of microscopic degrees of freedom has the greatest potential to explain
the appearance of {the inverse-catalysis effect.}
In particular, the dependence of the running coupling on the magnetic field is expected to play a
prominent role in a dynamical study of chiral symmetry breaking~\cite{Farias:2014eca,Ferrer:2014qka}. 

In this work, we analyze the origin of the inverse-catalysis effect in finite-temperature QCD. To this end, we discuss 
the chiral quark dynamics in the presence of an external magnetic field 
by studying the underlying fixed-point structure.
In particular, we shall point out that the dependence of the chiral dynamics on the external magnetic field
is governed by an intriguing interplay between the quark and gluon degrees of freedom.
However, we also emphasize that we are not aiming at quantitative precision {with our analysis but rather}
aim at revealing the mechanisms and the connection {underlying inverse catalysis} observed in lattice MC simulations and
magnetic catalysis being a well-established phenomenon in the context of condensed-matter systems.

{\it Formalism.--} The starting point for our analysis 
is the classical action~$S$ in 4$d$ Euclidean {space-time,
\be
S = \int \! d^4x\left\{ 
\frac{1}{4}F_{\mu\nu}F_{\mu\nu} 
+ \bar{\psi}\left(
{\rm i}\slashed{\partial} + \bar{g}\slashed{A} + \bar{e}{\mathcal A}\!\!\!\slash{}
\right)\psi
\right\}\,,\label{eq:qcd}
\ee
where~$\bar{g}$} is the bare gauge coupling and~$\bar{e}\equiv e$ denotes the electric charge. The non-Abelian fields~$A_{\mu}$
enter the definition of the field-strength tensor~$F_{\mu\nu}$ and
are associated with the gluon degrees of freedom. The external electrodynamic {potential is determined
by~${\mathcal A}_{\mu}=(0,0,Bx_1,0)$, where} the magnetic field~$B$ is assumed to be spatially and
temporarily constant. 
In the present work we shall moreover restrict ourselves to the case of two massless quark flavors.

The quark-gluon interaction in Eq.~\eqref{eq:qcd} induces quark self-interactions, e.~g. by two-gluon exchange. 
Chiral symmetry breaking is then ultimately triggered by the four-quark interactions approaching criticality, i.e. the 
associated couplings become relevant. In our study we shall only consider one particular four-quark 
interaction channel, namely the scalar-pseudoscalar
channel. This channel can be used to monitor spontaneous chiral symmetry breaking and 
has been found to be the most dominant one in Fierz-complete studies, both
{at zero~\cite{Gies:2005as,Mitter:2014wpa} and finite temperature~\cite{Braun:2005uj,*Braun:2006jd}. At} finite external magnetic field,
the number of interaction channels in a Fierz-complete basis increases considerably due to the strong explicit breaking
of Poincar{\'e} invariance~\cite{Scherer:2012nn,Ferrer:2013noa}. For our more qualitative analysis of the mechanisms underlying chiral
symmetry breaking, however, we shall drop these additional channels for simplicity.
\begin{figure}[t]
\includegraphics[scale=0.8]{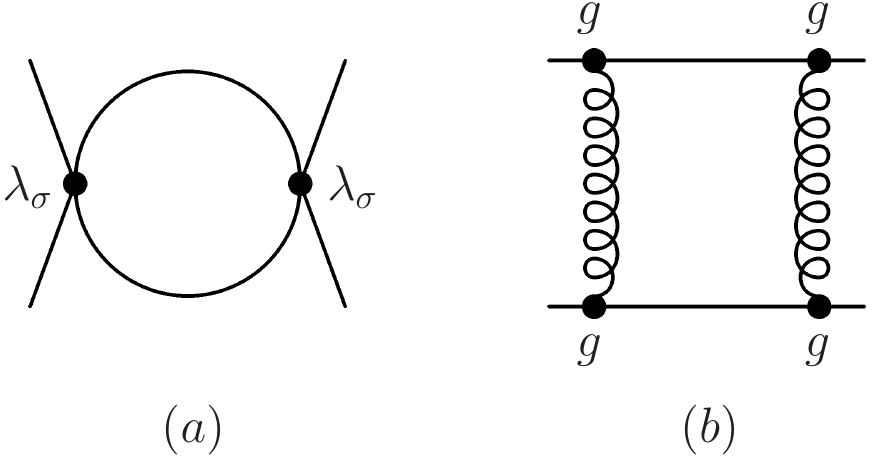}\hfill
\caption{The depicted Feynman diagrams labelled with (a) and (b) are
associated with the first and the third term on the right-hand side of Eq.~\eqref{eq:4psiflow}, respectively. The second
  term associated with the so-called triangle diagram is not shown, as it plays only a subleading role
  in our analysis.}
\label{fig:feynman}
\end{figure}

The renormalization-group (RG) $\beta$-function of the four-quark coupling~${\lambda}_{\sigma}$ associated 
with the scalar-pseudoscalar interaction channel is given {by
\be
\pat \lambda_{\sigma} = 2\lambda_{\sigma}^2 - l_{\lambda_{\sigma}^2}\,\lambda_{\sigma}^2 
- l_{\lambda_{\sigma}g^2}\,\lambda_{\sigma} g^2 - l_{g^4}\,g^4\,,
\label{eq:4psiflow}
\ee
where~$t=\ln(k/\Lambda)$ is} the so-called RG `time' with~$k$ being the RG scale and~$\Lambda$ being
an ultraviolet (UV) cutoff scale at which we fix the initial conditions, e.g. {the $\tau$-mass scale.}
The couplings~$\lambda_{\sigma}\sim k^2\bar{\lambda}_{\sigma}$ and~$g$ are dimensionless
and assumed to be suitably renormalized. The first {term on the right-hand side describes} simply the dimensional scaling
of the coupling. The quantities $l_{\lambda_{\sigma}^2} >0$, $l_{\lambda_{\sigma}g^2} >0$, {and~$l_{g^4} > 0$ are related}
to one-particle irreducible (1PI) Feynman diagrams, see Fig.~\ref{fig:feynman}, and
depend on the dimensionless magnetic field~$b=eB/k^2$, the dimensionless temperature~$\tau=T/k$ and combinatoric factors.
With respect to our numerical studies below, we add that Eq.~\eqref{eq:4psiflow} can also be derived from
nonperturbative flow equations in the limit of point-like interactions. Here,
we employ an RG equation for the quantum effective action~\cite{Wetterich:1992yh}. Our RG flow then takes into
account resummations of all diagram types shown in Fig.~\ref{fig:feynman}, including ladder diagrams.
The details of the Wilsonian
momentum-shell integrations are specified by the choice for a so-called regulator function in this case.
Here, we shall use {the exponential} regulator~\cite{Wetterich:1992yh,Jungnickel:1995fp}.

The initial condition for the four-quark coupling is fixed by the classical action~\eqref{eq:qcd} and
is therefore given by~$\lambda_{\sigma}=0$ at~$k=\Lambda\gg\Lambda_{\text{QCD}}$ 
in the perturbative high-momentum regime. Thus, the
quark self-interactions in our study are originally gluon-induced due to the term~$\sim g^4$
in Eq.~\eqref{eq:4psiflow} and the associated coupling~$\lambda_{\sigma}$ does
not represent a (free) parameter. The {\it only} parameter used in our numerical study below is
the value of the strong coupling $g^2/(4\pi)\approx 0.322$ at {the $\tau$-mass scale~$m_{\tau}\approx 1.78\,\text{GeV}$}
which determines the
physical scale~\cite{Bethke:2004uy}.

{\it Fixed-point analysis.--} {As indicated above,} chiral symmetry breaking is triggered by the four-quark interaction approaching criticality, i.e. 
the associated coupling~$\lambda_{\sigma}$ diverges in the {RG flow at a finite scale.} In fact, since the four-quark coupling~$\lambda_{\sigma}$ 
is the inverse mass parameter of a Ginzburg-Landau effective potential for the chiral order parameter, such a divergence 
indicates the {onset of chiral symmetry breaking, see}
Ref.~\cite{Braun:2011pp} for a review.

\begin{figure}[!t]
\includegraphics[scale=0.45,clip=true]{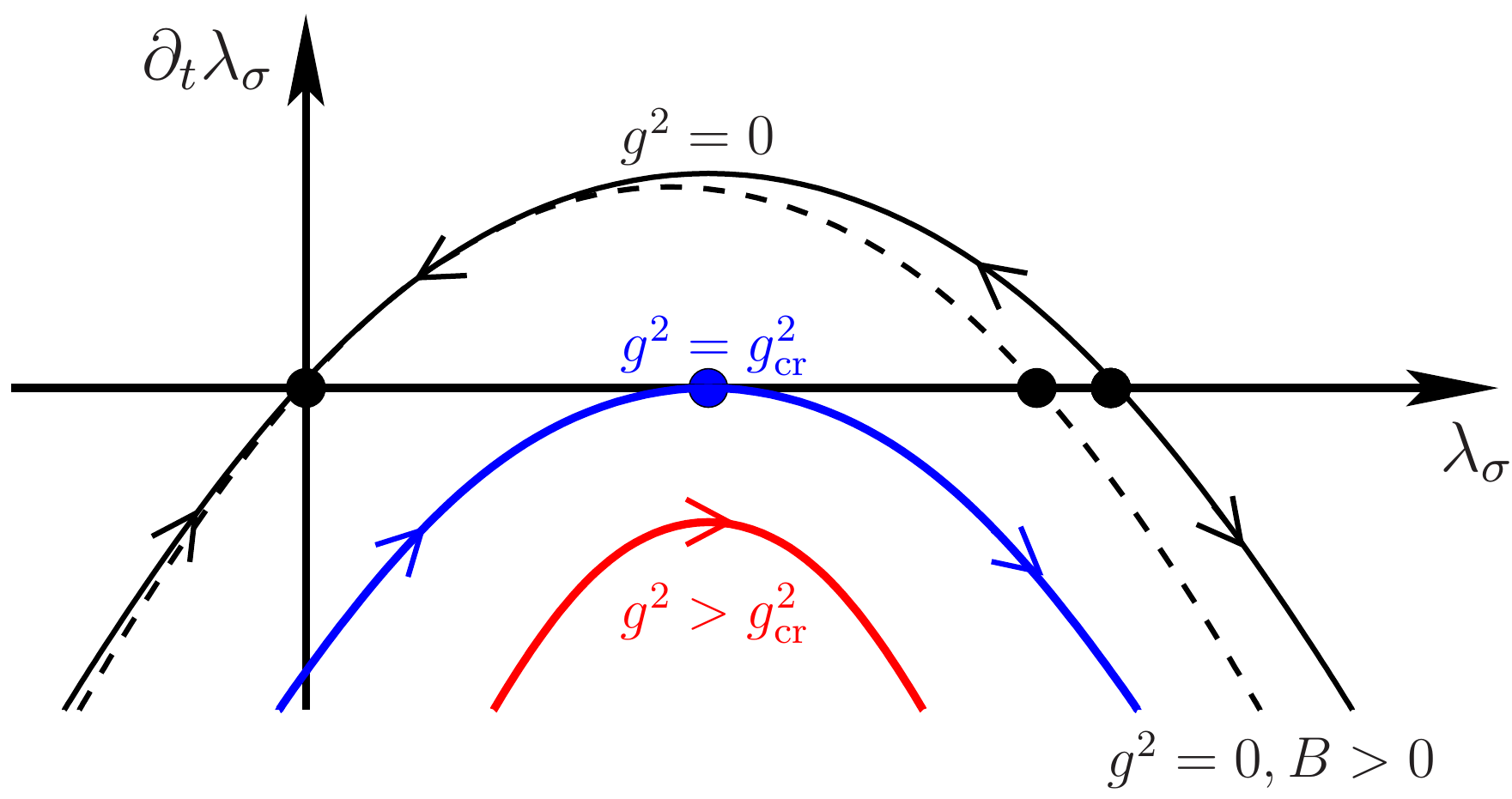} 
\caption{Sketch of the $\beta$ function of the four-quark
  interaction $\lambda_{\sigma}$ {at~$T=0=B$ (solid lines) as well} as for $g^2=0$ and $B>0$ (dashed line).
  The arrows indicate the direction of the flow towards the infrared.} 
\label{fig:parab}
\end{figure}

Let us now discuss spontaneous chiral symmetry breaking in QCD by analyzing 
the fixed-point structure of the flow equation~\eqref{eq:4psiflow}. We begin with the limit~$T=0$ and~$B=0$,
where the quantities $l_{\lambda_{\sigma}^2}$, $l_{\lambda_{\sigma}g^2}$, {and~$l_{g^4}$} 
are simply numbers. At weak gauge coupling, quark self-interactions
are exclusively generated by gluon exchange processes. For 
increasing gauge coupling, the two fixed points of the four-quark coupling then approach each other, see Fig.~\ref{fig:parab}.
Provided that the gauge coupling does not exceed a critical value,~$g^2 \leq g^2_{\text{cr}}$, the four-quark coupling
approaches a fixed point in the infrared and therefore remains finite. Thus, the system
stays in the chirally symmetric phase. However, if the gauge coupling exceeds the critical value~$g^2_{\text{cr}}$ at
some scale, then the four-quark coupling~$\lambda_{\sigma}$ is no longer bounded by fixed points. 
In fact,~$\lambda_{\sigma}$ grows rapidly and approaches a divergence at a finite scale, indicating
chiral symmetry breaking. 

At~$T>0$ and~$B=0$, the loop integrals parametrized by the quantities 
$l_{\lambda_{\sigma}^2}$, $l_{\lambda_{\sigma}g^2}$, {and~$l_{g^4}$}
become functions of the dimensionless temperature~$\tau=T/k$. For large scales~$k\gg T$, these functions
approach their zero-temperature values and the dynamics of the matter sector as measured by the fixed points
remains unchanged. For high temperatures (or small scales) $T\gg k$, on the other hand, the quarks
acquire a large thermal {mass and thus stronger interactions} are required to drive the quarks to
criticality. In other words, the critical value for the gauge coupling~$g_{\text{cr}}^2$ increases monotonically for increasing~$T/k$.
In fact, we have $l_{\lambda_{\sigma}^2}\to 0$, $l_{\lambda_{\sigma}g^2}\to 0$, {and~$l_{g^4}\to 0$} for~$T/k\to\infty$.
{This suggests chiral symmetry restoration at high temperatures.
This simple picture} of chiral symmetry breaking 
at zero and finite temperature has been put forward in 
Refs.~\cite{Gies:2005as,Braun:2005uj,Braun:2006jd}, {and underlies subsequent studies of 
the infrared properties of QCD~\cite{Mitter:2014wpa,Braun:2009gm,*Braun:2014ata}.}

We now discuss the fixed-point dynamics in the presence of an external magnetic field. To this end,
we first note that the gluons do not carry an electric charge
and are therefore not directly affected by the presence of an external magnetic field. On the other hand, the
quarks are electrically charged and their dynamics is indeed altered. 
More specifically, the magnetic field can be associated with a length scale~$\ell_{B}\sim (eB)^{-1/2}$. In the
presence of this scale, the quark fields experience Landau-level quantization which comes along
with a magnetic zero-mode (lowest Landau level) for the quark fields. This zero mode essentially governs the
dynamics of the quarks, in particular in the large-$b$ limit, {where~$b=eB/k^2$.}
In this work, we have computed the $B$-field dependence of the appearing loop integrals
in the lowest Landau-level (LLL) approximation and suitably amended them to approach the correct result
in the limit of large scales~$k$, i.e.~$b\to 0$. This ensures that the dynamics of the theory in the UV limit
remains unchanged. 

Let us begin our discussion of magnetic field effects
with the zero-temperature limit. Moreover, it is instructive to consider first the case of vanishing gauge coupling.
In this case, for increasing dimensionless magnetic field~$b=eB/k^2$, the {non-Gau\ss ian fixed point} is shifted to smaller
values which entails that the maximum of the $\beta$ function is also pushed to smaller values, see Fig.~\ref{fig:parab}.
This can be understood from the fact that the contribution from the quark loop increases monotonically with~$b$.
To be specific, we have~$l_{\lambda_{\sigma}^2}\sim b$ in the large-$b$ limit {which is well-known} from RG
studies of fermionic models~\cite{Scherer:2012nn,Fukushima}. In any case, 
the shift of the interacting {fixed point} towards the Gau\ss ian fixed point for finite~$b$ and~$g^2=0$ already
suggests that~$g^2_{\text{crit}}$ decreases
with increasing~$b$, at least in the large-$b$ limit. This is indeed the case since we find 
that the box diagram scales as~$l_{g^4}\sim b^{-1/2}$, resulting in~$g^2_{\text{cr}}\sim b^{-1/4}$ for large $b$, i.e.
in the deep infrared limit. The triangle diagram $l_{\lambda_{\sigma}g^2}\sim 1$ 
plays only a subleading role in our analysis. Note that the 
$b$-dependence of the quark loop in the LLL approximation is identical to
the $b$-scaling behavior of this diagram in the limit of asymptotically large magnetic fields. This is not the 
case for the triangle and the box diagram. Even in the LLL approximation, the dependence of these diagrams
on the magnetic field is more involved due to the internal gluon lines. The latter subtlety together with 
the observation that the term~$\sim g^4$ in Eq.~\eqref{eq:4psiflow} is suppressed for finite~$b$ is of great importance:
Lowering the scale~$k$ starting from a point in the UV regime with $g^2_{\text{cr}}(b)\approx g^2_{\text{cr}}(0)$, 
we find that the magnetic suppression of the box diagram yields an increase of~$g^2_{\text{cr}}$ before it reaches
a maximum and then approaches zero due to the magnetic enhancement of the quark loop in the infrared limit.

The scale dependence of $g^2_{\text{cr}}$ for a given finite temperature~$T$ and magnetic field~$B$ can
now be understood from our discussion above. For large scales~$k\gg T$ and~$k\gg B$, the chiral critical coupling 
$g^2_{\text{cr}}$ approaches a finite constant
value, $g^2_{\text{cr}}(b,\tau)\!\to\! g^2_{\text{cr}}(0,0)$.
Starting in the UV limit and lowering the scale~$k$, we find that~$g^2_{\text{cr}}$ increases. Most importantly, 
$g^2_{\text{cr}}$ becomes even larger than in the case~$B\!=\!0$ for the same temperature due to the magnetic suppression
of the box diagram. It is this increase of~$g^2_{\text{cr}}$ 
on intermediate scales at finite~$B$ which 
{favors inverse catalysis} over catalysis for weak magnetic fields.

{\it Running gauge coupling.--}  
Our fixed-point analysis allows us to trace the question of the onset of chiral symmetry 
breaking back to the strength of the coupling~$g^2$ relative to the critical coupling~$g^2_{\text{cr}}$.
Therefore an actual determination of the QCD ground-state properties with 
respect to chiral symmetry requires information about the RG running of the strong coupling.
To this end, we employ the results for the gauge coupling at zero and finite temperature
from Refs.~\cite{Gies:2002af,Braun:2005uj,Braun:2006jd} and restrict ourselves to Feynman 
gauge for simplicity.
At zero temperature, the running of~$g^2$ agrees well with perturbation theory for small
coupling. In the infrared limit, on the other hand, the gauge coupling
assumes a finite value associated with an a non-Gau\ss ian fixed point.
At finite temperature, the behavior of the coupling at large scales~$k\gg T$ remains unaffected
and still agrees well with the perturbative running in the zero-temperature limit.
In the infrared limit, however, the running of the coupling has been found to be qualitatively distinct from the 
zero-temperature case~\cite{Braun:2005uj,Braun:2006jd}. 
In fact, the coupling decreases linearly with the scale~$k$ according to~$g^2 \sim k/T$ and eventually tends to zero. 
In this regime, the flow of the 
running coupling is solely driven by the gluonic thermal zero-mode associated
with the spatial 3$d$ Yang-Mills theory. Thus, the infrared behavior of the coupling in 4$d$
is directly related to the infrared behavior of the coupling in the underlying 3$d$ 
theory: $g^2(k\ll T)\sim g^2_{3d,\ast}\, k/T$ with~$g^2_{3d,\ast}\sim {\mathcal O}(1)$, see Ref.~\cite{Braun:2005uj,Braun:2006jd}.
Moreover, we note that the running coupling
is bounded from above on all scales and the maximum value of the coupling decreases with increasing
temperature.
\begin{figure}[t]
\includegraphics[scale=0.7]{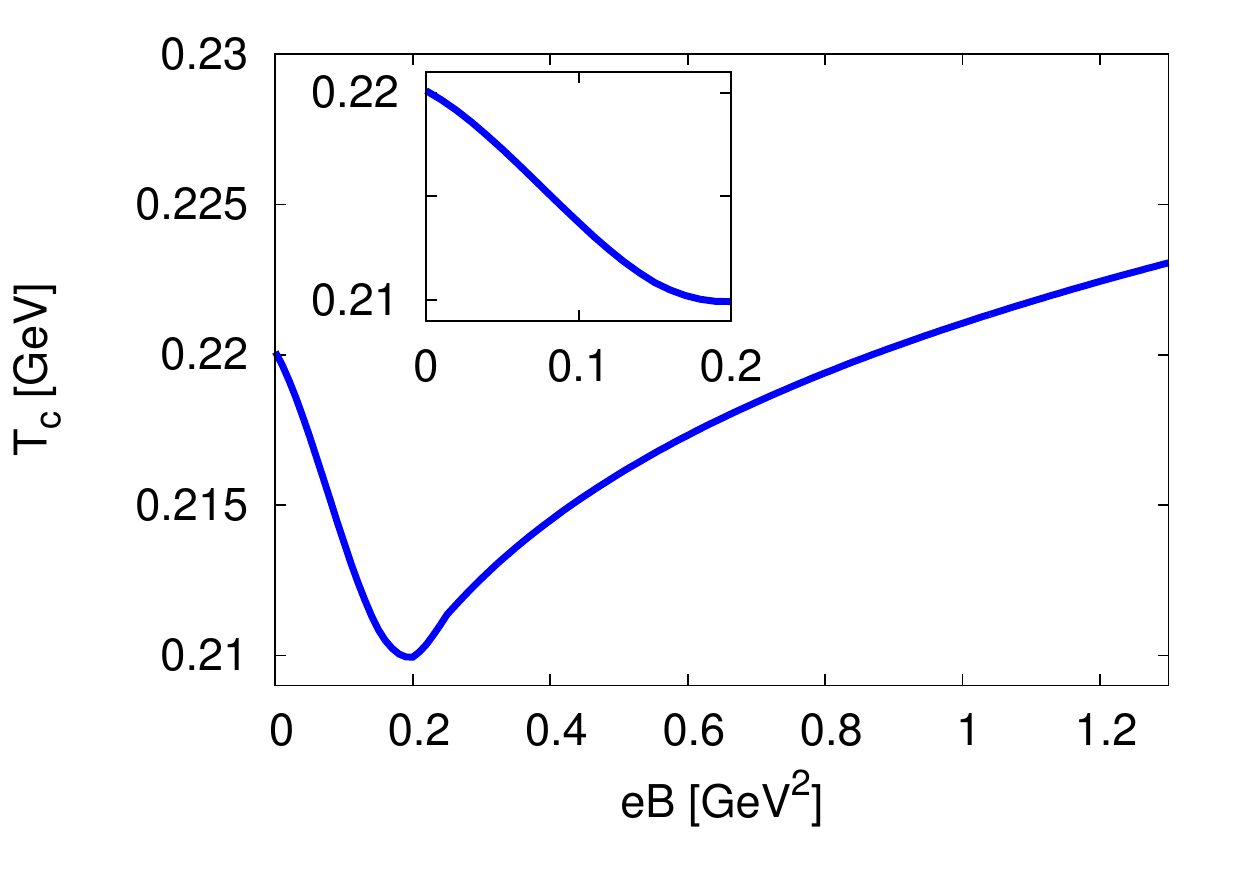}
\caption{Chiral critical temperature
$\Tc$ as a function of~$eB$.
\label{fig:tccrit} }
\end{figure}

At finite magnetic field, the gauge coupling~$g^2$ also depends on~$b=eB/k^2$ due
to the quark-gluon coupling. 
To better understand the effect of the magnetic field on~$g^2$,
we consider the associated $\beta$ function: $\pat g^2 = \eta_{g^2}g^2 = (\eta_A + \eta_q)g^2$, 
which we have conveniently split into a purely gluonic contribution~$\eta_A$ and a contribution~$\eta_q$ containing
all contributions from diagrams with internal quark lines. For example, at the one-loop level, $\eta_q \sim g^2$
{is structurally identical} to the vacuum polarisation tensor in Quantum Electrodynamics (QED). In the limit of large
(dimensionless) magnetic fields, $\eta_q$ is primarily determined by the dynamics of the magnetic zero mode of the quarks,
resulting in~$\eta_q \sim b$ at zero temperature as in QED~\cite{Karbstein:2011ja}.
Thus, the quark contribution~$\eta_q$ to the running coupling is magnetically
enhanced compared to the gluon contribution~$\eta_A$. This results in {a decrease of the} coupling in the infrared 
limit,~$g^2\! \to\! 0$ for $k\!\to\! 0$. Loosely speaking, {the magnetic field acts as an increase}
of quark flavors towards the infrared limit, such that the gauge coupling is attracted by the Gau\ss ian fixed point.
In contrast to the finite-temperature case at~$B=0$, the behavior of the gauge coupling
in the infrared can now not be straightforwardly related to the dynamics of a dimensionally reduced gauge theory:
Whereas the quark fields experience a dimensional reduction from four to two
dimensions for strong magnetic fields, the gluon fields are not directly affected
and therefore are not dimensionally reduced in the infrared limit. 

In the presence of a magnetic field at finite temperature, the running of the gauge coupling in the (deep) infrared
is still governed by the underlying 3$d$ Yang-Mills theory since the quarks decouple eventually from the flow 
due to their thermal Matsubara masses. In any case, for
our numerical computation of the chiral critical temperature~$\Tc$ as a function of the magnetic field~$B$, we 
employed the results for the running gauge coupling from Ref.~\cite{Braun:2005uj,Braun:2006jd} 
and used only the $B$-field dependent one-loop expression for~$\eta_q$
in the LLL approximation to suitably amend the quark contribution~$\eta_q$ to~$\eta_{g^2}$ in the large-$b$ limit;
details will be presented elsewhere~\cite{BMR}.

{\it Magnetic phase diagram.--} Using our numerical results for the scale-dependence of the chiral critical coupling~$g^2_{\text{cr}}$ and
the running gauge coupling~$g^2$, we can now compute the dependence of the chiral
phase transition temperature~$\Tc$ on the magnetic field~$B$. To be specific, we 
estimate $\Tc$ as the lowest temperature for which no intersection
point between $g^2$ and $g^2_{\text{cr}}$ as a function of~$k$ occurs for a given value of the magnetic field~$B$. 
In Fig.~\ref{fig:tccrit}, we show~$\Tc$ as a function of~$eB$.
For~$B=0$, we find~$\Tc \approx 220\,\text{MeV}$. The difference to the accepted value for
the critical temperature from lattice QCD simulations~\cite{Aoki:2006we,*Aoki:2006br,*Cheng:2007jq} can be traced back to the fact that we did not
consider a Fierz-complete set of four-quark interactions~\cite{Braun:2005uj,Braun:2006jd}. In any
case, increasing the magnetic field~$B$, we find that the critical temperature~$\Tc$ decreases as also
observed in lattice MC simulations~\cite{Bali:2011qj,Bali:2012zg}. We note that this decrease of~$\Tc$
persists, even if we consider a $B$-independent running coupling. 
Thus, this decrease in the phase transition temperature can be traced back to the 
dynamics in the matter sector as discussed above in terms of our fixed-point analysis.
The {effective decrease} of the gauge coupling at finite magnetic field only intensifies the inverse-catalysis effect
in our analysis. 

Increasing the magnetic field further, we observe that~$\Tc(eB)$ assumes a minimum at~$eB\approx 0.2\,\text{GeV}^2$ 
and then increases for larger values of~$eB$. 
{Indications for such an increase at strong magnetic fields are also seen in a DSE study~\cite{MP}.
The catalysis effect} for large~$eB$ can be traced back
to the fact that the RG running of the four-quark interaction is mainly driven by the quark loop at strong magnetic fields.
This results in a decrease of the critical coupling~$g^2_{\text{cr}}$ and, in turn, in an increase of the critical temperature.
Thus, the well-established magnetic catalysis effect in fermionic theories, which is simply driven by the fermion loop in Fig.~\ref{fig:feynman}~(a), 
sets in ``delayed" due to the non-trivial quark-gluon dynamics in the matter sector.

{\it Conclusions. --} We have {computed} the phase diagram of QCD in the
plane of temperature and magnetic field. Our results {confirm the existence of the inverse-catalysis effect.}
Compared to lattice MC results for $2+1$ {(massive) quark flavors}~\cite{Bali:2011qj,Bali:2012zg}, our 
RG analysis predicts a smaller regime in which inverse catalysis occurs. Clearly, our
simple study based on a single four-quark channel cannot be expected to
be quantitative. Still, it appears worthwhile to study the scaling of the size of the inverse-catalysis regime
when the number of quark flavors is increased. 
In any case, for large magnetic fields, we observe magnetic catalysis. Our fixed-point analysis reveals a simple mechanism {for inverse magnetic catalysis at}
weak magnetic fields and, at the same time, explains the dynamics underlying the observed magnetic catalysis
at strong magnetic fields. In this respect, the {observed} ``delayed" magnetic catalysis can be viewed
as a testable prediction for future {lattice MC studies.} Moreover, our simple analysis represents
a promising starting point for phenomenological applications, such as the microscopically guided 
improvement of well-established
QCD models.

\medskip

{\it Acknowledgments.--~} The authors thank H.~Gies, F.~Karbstein, J.~M. Pawlowski, D.~Roscher, and D.~D.~Scherer
for useful discussions. Moreover, the authors are grateful to H.~Gies and J.~M. Pawlowski for comments
on the manuscript. J.B. and S.R. acknowledge support
by HIC for FAIR within the LOEWE program of the State of Hesse as well as 
by the DFG under grant SFB 634.

%
\bibliography{references}

\begin{thebibliography}{48}%
\makeatletter
\providecommand \@ifxundefined [1]{%
 \@ifx{#1\undefined}
}%
\providecommand \@ifnum [1]{%
 \ifnum #1\expandafter \@firstoftwo
 \else \expandafter \@secondoftwo
 \fi
}%
\providecommand \@ifx [1]{%
 \ifx #1\expandafter \@firstoftwo
 \else \expandafter \@secondoftwo
 \fi
}%
\providecommand \natexlab [1]{#1}%
\providecommand \enquote  [1]{``#1''}%
\providecommand \bibnamefont  [1]{#1}%
\providecommand \bibfnamefont [1]{#1}%
\providecommand \citenamefont [1]{#1}%
\providecommand \href@noop [0]{\@secondoftwo}%
\providecommand \href [0]{\begingroup \@sanitize@url \@href}%
\providecommand \@href[1]{\@@startlink{#1}\@@href}%
\providecommand \@@href[1]{\endgroup#1\@@endlink}%
\providecommand \@sanitize@url [0]{\catcode `\\12\catcode `\$12\catcode
  `\&12\catcode `\#12\catcode `\^12\catcode `\_12\catcode `\%12\relax}%
\providecommand \@@startlink[1]{}%
\providecommand \@@endlink[0]{}%
\providecommand \url  [0]{\begingroup\@sanitize@url \@url }%
\providecommand \@url [1]{\endgroup\@href {#1}{\urlprefix }}%
\providecommand \urlprefix  [0]{URL }%
\providecommand \Eprint [0]{\href }%
\providecommand \doibase [0]{http://dx.doi.org/}%
\providecommand \selectlanguage [0]{\@gobble}%
\providecommand \bibinfo  [0]{\@secondoftwo}%
\providecommand \bibfield  [0]{\@secondoftwo}%
\providecommand \translation [1]{[#1]}%
\providecommand \BibitemOpen [0]{}%
\providecommand \bibitemStop [0]{}%
\providecommand \bibitemNoStop [0]{.\EOS\space}%
\providecommand \EOS [0]{\spacefactor3000\relax}%
\providecommand \BibitemShut  [1]{\csname bibitem#1\endcsname}%
\let\auto@bib@innerbib\@empty
\bibitem [{\citenamefont {Gusynin}\ \emph
  {et~al.}(1995{\natexlab{a}})\citenamefont {Gusynin}, \citenamefont
  {Miransky},\ and\ \citenamefont {Shovkovy}}]{Gusynin:1994va}%
  \BibitemOpen
  \bibfield  {author} {\bibinfo {author} {\bibfnamefont {V.}~\bibnamefont
  {Gusynin}}, \bibinfo {author} {\bibfnamefont {V.}~\bibnamefont {Miransky}}, \
  and\ \bibinfo {author} {\bibfnamefont {I.}~\bibnamefont {Shovkovy}},\ }\href
  {\doibase 10.1103/PhysRevD.52.4718} {\bibfield  {journal} {\bibinfo
  {journal} {Phys.Rev.}\ }\textbf {\bibinfo {volume} {D52}},\ \bibinfo {pages}
  {4718} (\bibinfo {year} {1995}{\natexlab{a}})}\BibitemShut {NoStop}%
\bibitem [{\citenamefont {{Gusynin}}\ and\ \citenamefont
  {{Sharapov}}(2005)}]{Gusynin}%
  \BibitemOpen
  \bibfield  {author} {\bibinfo {author} {\bibfnamefont {V.~P.}\ \bibnamefont
  {{Gusynin}}}\ and\ \bibinfo {author} {\bibfnamefont {S.~G.}\ \bibnamefont
  {{Sharapov}}},\ }\href {\doibase 10.1103/PhysRevLett.95.146801} {\bibfield
  {journal} {\bibinfo  {journal} {Phys.Rev.Lett.}\ }\textbf {\bibinfo {volume}
  {95}},\ \bibinfo {eid} {146801} (\bibinfo {year} {2005})}\BibitemShut
  {NoStop}%
\bibitem [{\citenamefont {Kharzeev}\ \emph {et~al.}(2008)\citenamefont
  {Kharzeev}, \citenamefont {McLerran},\ and\ \citenamefont
  {Warringa}}]{Kharzeev:2007jp}%
  \BibitemOpen
  \bibfield  {author} {\bibinfo {author} {\bibfnamefont {D.~E.}\ \bibnamefont
  {Kharzeev}}, \bibinfo {author} {\bibfnamefont {L.~D.}\ \bibnamefont
  {McLerran}}, \ and\ \bibinfo {author} {\bibfnamefont {H.~J.}\ \bibnamefont
  {Warringa}},\ }\href {\doibase 10.1016/j.nuclphysa.2008.02.298} {\bibfield
  {journal} {\bibinfo  {journal} {Nucl.Phys.}\ }\textbf {\bibinfo {volume}
  {A803}},\ \bibinfo {pages} {227} (\bibinfo {year} {2008})}\BibitemShut
  {NoStop}%
\bibitem [{\citenamefont {Skokov}\ \emph {et~al.}(2009)\citenamefont {Skokov},
  \citenamefont {Illarionov},\ and\ \citenamefont {Toneev}}]{Skokov:2009qp}%
  \BibitemOpen
  \bibfield  {author} {\bibinfo {author} {\bibfnamefont {V.}~\bibnamefont
  {Skokov}}, \bibinfo {author} {\bibfnamefont {A.~Y.}\ \bibnamefont
  {Illarionov}}, \ and\ \bibinfo {author} {\bibfnamefont {V.}~\bibnamefont
  {Toneev}},\ }\href {\doibase 10.1142/S0217751X09047570} {\bibfield  {journal}
  {\bibinfo  {journal} {Int.J.Mod.Phys.}\ }\textbf {\bibinfo {volume} {A24}},\
  \bibinfo {pages} {5925} (\bibinfo {year} {2009})}\BibitemShut {NoStop}%
\bibitem [{\citenamefont {Duncan}\ and\ \citenamefont
  {Thompson}(1992)}]{Duncan}%
  \BibitemOpen
  \bibfield  {author} {\bibinfo {author} {\bibfnamefont {R.~C.}\ \bibnamefont
  {Duncan}}\ and\ \bibinfo {author} {\bibfnamefont {C.}~\bibnamefont
  {Thompson}},\ }\href@noop {} {\bibfield  {journal} {\bibinfo  {journal}
  {Astrophys. J.}\ }\textbf {\bibinfo {volume} {392}},\ \bibinfo {pages} {L9}
  (\bibinfo {year} {1992})}\BibitemShut {NoStop}%
\bibitem [{\citenamefont {Vachaspati}(1991)}]{Vachaspati}%
  \BibitemOpen
  \bibfield  {author} {\bibinfo {author} {\bibfnamefont {T.}~\bibnamefont
  {Vachaspati}},\ }\href@noop {} {\bibfield  {journal} {\bibinfo  {journal}
  {Phys. Lett.}\ }\textbf {\bibinfo {volume} {B265}},\ \bibinfo {pages} {258}
  (\bibinfo {year} {1991})}\BibitemShut {NoStop}%
\bibitem [{\citenamefont {Bali}\ \emph
  {et~al.}(2012{\natexlab{a}})\citenamefont {Bali}, \citenamefont {Bruckmann},
  \citenamefont {Endrodi}, \citenamefont {Fodor}, \citenamefont {Katz} \emph
  {et~al.}}]{Bali:2011qj}%
  \BibitemOpen
  \bibfield  {author} {\bibinfo {author} {\bibfnamefont {G.}~\bibnamefont
  {Bali}}, \bibinfo {author} {\bibfnamefont {F.}~\bibnamefont {Bruckmann}},
  \bibinfo {author} {\bibfnamefont {G.}~\bibnamefont {Endrodi}}, \bibinfo
  {author} {\bibfnamefont {Z.}~\bibnamefont {Fodor}}, \bibinfo {author}
  {\bibfnamefont {S.}~\bibnamefont {Katz}},  \emph {et~al.},\ }\href {\doibase
  10.1007/JHEP02(2012)044} {\bibfield  {journal} {\bibinfo  {journal} {JHEP}\
  }\textbf {\bibinfo {volume} {1202}},\ \bibinfo {pages} {044} (\bibinfo {year}
  {2012}{\natexlab{a}})}\BibitemShut {NoStop}%
\bibitem [{\citenamefont {Bali}\ \emph
  {et~al.}(2012{\natexlab{b}})\citenamefont {Bali}, \citenamefont {Bruckmann},
  \citenamefont {Endrodi}, \citenamefont {Fodor}, \citenamefont {Katz} \emph
  {et~al.}}]{Bali:2012zg}%
  \BibitemOpen
  \bibfield  {author} {\bibinfo {author} {\bibfnamefont {G.}~\bibnamefont
  {Bali}}, \bibinfo {author} {\bibfnamefont {F.}~\bibnamefont {Bruckmann}},
  \bibinfo {author} {\bibfnamefont {G.}~\bibnamefont {Endrodi}}, \bibinfo
  {author} {\bibfnamefont {Z.}~\bibnamefont {Fodor}}, \bibinfo {author}
  {\bibfnamefont {S.}~\bibnamefont {Katz}},  \emph {et~al.},\ }\href {\doibase
  10.1103/PhysRevD.86.071502} {\bibfield  {journal} {\bibinfo  {journal}
  {Phys.Rev.}\ }\textbf {\bibinfo {volume} {D86}},\ \bibinfo {pages} {071502}
  (\bibinfo {year} {2012}{\natexlab{b}})}\BibitemShut {NoStop}%
\bibitem [{\citenamefont {Gusynin}\ \emph
  {et~al.}(1995{\natexlab{b}})\citenamefont {Gusynin}, \citenamefont
  {Miransky},\ and\ \citenamefont {Shovkovy}}]{Gusynin:1994xp}%
  \BibitemOpen
  \bibfield  {author} {\bibinfo {author} {\bibfnamefont {V.}~\bibnamefont
  {Gusynin}}, \bibinfo {author} {\bibfnamefont {V.}~\bibnamefont {Miransky}}, \
  and\ \bibinfo {author} {\bibfnamefont {I.}~\bibnamefont {Shovkovy}},\ }\href
  {\doibase 10.1016/0370-2693(95)00232-A} {\bibfield  {journal} {\bibinfo
  {journal} {Phys.Lett.}\ }\textbf {\bibinfo {volume} {B349}},\ \bibinfo
  {pages} {477} (\bibinfo {year} {1995}{\natexlab{b}})}\BibitemShut {NoStop}%
\bibitem [{\citenamefont {Gusynin}\ \emph {et~al.}(1996)\citenamefont
  {Gusynin}, \citenamefont {Miransky},\ and\ \citenamefont
  {Shovkovy}}]{Gusynin:1995nb}%
  \BibitemOpen
  \bibfield  {author} {\bibinfo {author} {\bibfnamefont {V.}~\bibnamefont
  {Gusynin}}, \bibinfo {author} {\bibfnamefont {V.}~\bibnamefont {Miransky}}, \
  and\ \bibinfo {author} {\bibfnamefont {I.}~\bibnamefont {Shovkovy}},\ }\href
  {\doibase 10.1016/0550-3213(96)00021-1} {\bibfield  {journal} {\bibinfo
  {journal} {Nucl.Phys.}\ }\textbf {\bibinfo {volume} {B462}},\ \bibinfo
  {pages} {249} (\bibinfo {year} {1996})}\BibitemShut {NoStop}%
\bibitem [{\citenamefont {Nambu}\ and\ \citenamefont
  {Jona-Lasinio}(1961{\natexlab{a}})}]{Nambu:1961tp}%
  \BibitemOpen
  \bibfield  {author} {\bibinfo {author} {\bibfnamefont {Y.}~\bibnamefont
  {Nambu}}\ and\ \bibinfo {author} {\bibfnamefont {G.}~\bibnamefont
  {Jona-Lasinio}},\ }\href {\doibase 10.1103/PhysRev.122.345} {\bibfield
  {journal} {\bibinfo  {journal} {Phys. Rev.}\ }\textbf {\bibinfo {volume}
  {122}},\ \bibinfo {pages} {345} (\bibinfo {year}
  {1961}{\natexlab{a}})}\BibitemShut {NoStop}%
\bibitem [{\citenamefont {Nambu}\ and\ \citenamefont
  {Jona-Lasinio}(1961{\natexlab{b}})}]{Nambu:1961fr}%
  \BibitemOpen
  \bibfield  {author} {\bibinfo {author} {\bibfnamefont {Y.}~\bibnamefont
  {Nambu}}\ and\ \bibinfo {author} {\bibfnamefont {G.}~\bibnamefont
  {Jona-Lasinio}},\ }\href {\doibase 10.1103/PhysRev.124.246} {\bibfield
  {journal} {\bibinfo  {journal} {Phys. Rev.}\ }\textbf {\bibinfo {volume}
  {124}},\ \bibinfo {pages} {246} (\bibinfo {year}
  {1961}{\natexlab{b}})}\BibitemShut {NoStop}%
\bibitem [{\citenamefont {Osipov}\ \emph {et~al.}(2007)\citenamefont {Osipov},
  \citenamefont {Hiller}, \citenamefont {Blin},\ and\ \citenamefont
  {da~Providencia}}]{Osipov:2007je}%
  \BibitemOpen
  \bibfield  {author} {\bibinfo {author} {\bibfnamefont {A.}~\bibnamefont
  {Osipov}}, \bibinfo {author} {\bibfnamefont {B.}~\bibnamefont {Hiller}},
  \bibinfo {author} {\bibfnamefont {A.}~\bibnamefont {Blin}}, \ and\ \bibinfo
  {author} {\bibfnamefont {J.}~\bibnamefont {da~Providencia}},\ }\href
  {\doibase 10.1016/j.physletb.2007.05.033} {\bibfield  {journal} {\bibinfo
  {journal} {Phys.Lett.}\ }\textbf {\bibinfo {volume} {B650}},\ \bibinfo
  {pages} {262} (\bibinfo {year} {2007})}\BibitemShut {NoStop}%
\bibitem [{\citenamefont {Fraga}\ and\ \citenamefont
  {Mizher}(2008)}]{Fraga:2008qn}%
  \BibitemOpen
  \bibfield  {author} {\bibinfo {author} {\bibfnamefont {E.~S.}\ \bibnamefont
  {Fraga}}\ and\ \bibinfo {author} {\bibfnamefont {A.~J.}\ \bibnamefont
  {Mizher}},\ }\href {\doibase 10.1103/PhysRevD.78.025016} {\bibfield
  {journal} {\bibinfo  {journal} {Phys.Rev.}\ }\textbf {\bibinfo {volume}
  {D78}},\ \bibinfo {pages} {025016} (\bibinfo {year} {2008})}\BibitemShut
  {NoStop}%
\bibitem [{\citenamefont {Shovkovy}(2013)}]{Shovkovy:2012zn}%
  \BibitemOpen
  \bibfield  {author} {\bibinfo {author} {\bibfnamefont {I.~A.}\ \bibnamefont
  {Shovkovy}},\ }\href {\doibase 10.1007/978-3-642-37305-3_2} {\bibfield
  {journal} {\bibinfo  {journal} {Lect.Notes Phys.}\ }\textbf {\bibinfo
  {volume} {871}},\ \bibinfo {pages} {13} (\bibinfo {year} {2013})}\BibitemShut
  {NoStop}%
\bibitem [{\citenamefont {Andersen}\ \emph {et~al.}()\citenamefont {Andersen},
  \citenamefont {Naylor},\ and\ \citenamefont {Tranberg}}]{Andersen:2014xxa}%
  \BibitemOpen
  \bibfield  {author} {\bibinfo {author} {\bibfnamefont {J.~O.}\ \bibnamefont
  {Andersen}}, \bibinfo {author} {\bibfnamefont {W.~R.}\ \bibnamefont
  {Naylor}}, \ and\ \bibinfo {author} {\bibfnamefont {A.}~\bibnamefont
  {Tranberg}},\ }\href@noop {} {\ }\Eprint {http://arxiv.org/abs/1411.7176}
  {arXiv:1411.7176 [hep-ph]} \BibitemShut {NoStop}%
\bibitem [{\citenamefont {Mizher}\ \emph {et~al.}(2010)\citenamefont {Mizher},
  \citenamefont {Chernodub},\ and\ \citenamefont {Fraga}}]{Mizher:2010zb}%
  \BibitemOpen
  \bibfield  {author} {\bibinfo {author} {\bibfnamefont {A.~J.}\ \bibnamefont
  {Mizher}}, \bibinfo {author} {\bibfnamefont {M.}~\bibnamefont {Chernodub}}, \
  and\ \bibinfo {author} {\bibfnamefont {E.~S.}\ \bibnamefont {Fraga}},\ }\href
  {\doibase 10.1103/PhysRevD.82.105016} {\bibfield  {journal} {\bibinfo
  {journal} {Phys.Rev.}\ }\textbf {\bibinfo {volume} {D82}},\ \bibinfo {pages}
  {105016} (\bibinfo {year} {2010})}\BibitemShut {NoStop}%
\bibitem [{\citenamefont {Gatto}\ and\ \citenamefont
  {Ruggieri}(2010)}]{Gatto:2010qs}%
  \BibitemOpen
  \bibfield  {author} {\bibinfo {author} {\bibfnamefont {R.}~\bibnamefont
  {Gatto}}\ and\ \bibinfo {author} {\bibfnamefont {M.}~\bibnamefont
  {Ruggieri}},\ }\href {\doibase 10.1103/PhysRevD.82.054027} {\bibfield
  {journal} {\bibinfo  {journal} {Phys.Rev.}\ }\textbf {\bibinfo {volume}
  {D82}},\ \bibinfo {pages} {054027} (\bibinfo {year} {2010})}\BibitemShut
  {NoStop}%
\bibitem [{\citenamefont {Gatto}\ and\ \citenamefont
  {Ruggieri}(2011)}]{Gatto:2010pt}%
  \BibitemOpen
  \bibfield  {author} {\bibinfo {author} {\bibfnamefont {R.}~\bibnamefont
  {Gatto}}\ and\ \bibinfo {author} {\bibfnamefont {M.}~\bibnamefont
  {Ruggieri}},\ }\href {\doibase 10.1103/PhysRevD.83.034016} {\bibfield
  {journal} {\bibinfo  {journal} {Phys.Rev.}\ }\textbf {\bibinfo {volume}
  {D83}},\ \bibinfo {pages} {034016} (\bibinfo {year} {2011})}\BibitemShut
  {NoStop}%
\bibitem [{\citenamefont {Kashiwa}(2011)}]{Kashiwa:2011js}%
  \BibitemOpen
  \bibfield  {author} {\bibinfo {author} {\bibfnamefont {K.}~\bibnamefont
  {Kashiwa}},\ }\href {\doibase 10.1103/PhysRevD.83.117901} {\bibfield
  {journal} {\bibinfo  {journal} {Phys.Rev.}\ }\textbf {\bibinfo {volume}
  {D83}},\ \bibinfo {pages} {117901} (\bibinfo {year} {2011})}\BibitemShut
  {NoStop}%
\bibitem [{\citenamefont {Fukushima}\ and\ \citenamefont
  {Pawlowski}(2012)}]{Fukushima}%
  \BibitemOpen
  \bibfield  {author} {\bibinfo {author} {\bibfnamefont {K.}~\bibnamefont
  {Fukushima}}\ and\ \bibinfo {author} {\bibfnamefont {J.~M.}\ \bibnamefont
  {Pawlowski}},\ }\href {\doibase 10.1103/PhysRevD.86.076013} {\bibfield
  {journal} {\bibinfo  {journal} {Phys.Rev.}\ }\textbf {\bibinfo {volume}
  {D86}},\ \bibinfo {pages} {076013} (\bibinfo {year} {2012})}\BibitemShut
  {NoStop}%
\bibitem [{\citenamefont {Skokov}(2012)}]{Skokov:2011ib}%
  \BibitemOpen
  \bibfield  {author} {\bibinfo {author} {\bibfnamefont {V.}~\bibnamefont
  {Skokov}},\ }\href {\doibase 10.1103/PhysRevD.85.034026} {\bibfield
  {journal} {\bibinfo  {journal} {Phys.Rev.}\ }\textbf {\bibinfo {volume}
  {D85}},\ \bibinfo {pages} {034026} (\bibinfo {year} {2012})}\BibitemShut
  {NoStop}%
\bibitem [{\citenamefont {Kamikado}\ and\ \citenamefont
  {Kanazawa}(2014)}]{Kamikado:2013pya}%
  \BibitemOpen
  \bibfield  {author} {\bibinfo {author} {\bibfnamefont {K.}~\bibnamefont
  {Kamikado}}\ and\ \bibinfo {author} {\bibfnamefont {T.}~\bibnamefont
  {Kanazawa}},\ }\href {\doibase 10.1007/JHEP03(2014)009} {\bibfield  {journal}
  {\bibinfo  {journal} {JHEP}\ }\textbf {\bibinfo {volume} {1403}},\ \bibinfo
  {pages} {009} (\bibinfo {year} {2014})}\BibitemShut {NoStop}%
\bibitem [{\citenamefont {Kamikado}\ and\ \citenamefont
  {Kanazawa}()}]{Kamikado:2014bua}%
  \BibitemOpen
  \bibfield  {author} {\bibinfo {author} {\bibfnamefont {K.}~\bibnamefont
  {Kamikado}}\ and\ \bibinfo {author} {\bibfnamefont {T.}~\bibnamefont
  {Kanazawa}},\ }\href@noop {} {\ }\Eprint {http://arxiv.org/abs/1410.6253}
  {arXiv:1410.6253 [hep-ph]} \BibitemShut {NoStop}%
\bibitem [{\citenamefont {Mueller}\ \emph {et~al.}(2014)\citenamefont
  {Mueller}, \citenamefont {Bonnet},\ and\ \citenamefont
  {Fischer}}]{Mueller:2014tea}%
  \BibitemOpen
  \bibfield  {author} {\bibinfo {author} {\bibfnamefont {N.}~\bibnamefont
  {Mueller}}, \bibinfo {author} {\bibfnamefont {J.~A.}\ \bibnamefont {Bonnet}},
  \ and\ \bibinfo {author} {\bibfnamefont {C.~S.}\ \bibnamefont {Fischer}},\
  }\href {\doibase 10.1103/PhysRevD.89.094023} {\bibfield  {journal} {\bibinfo
  {journal} {Phys.Rev.}\ }\textbf {\bibinfo {volume} {D89}},\ \bibinfo {pages}
  {094023} (\bibinfo {year} {2014})}\BibitemShut {NoStop}%
\bibitem [{\citenamefont {Fraga}\ \emph {et~al.}(2014)\citenamefont {Fraga},
  \citenamefont {Mintz},\ and\ \citenamefont
  {Schaffner-Bielich}}]{Fraga:2013ova}%
  \BibitemOpen
  \bibfield  {author} {\bibinfo {author} {\bibfnamefont {E.}~\bibnamefont
  {Fraga}}, \bibinfo {author} {\bibfnamefont {B.}~\bibnamefont {Mintz}}, \ and\
  \bibinfo {author} {\bibfnamefont {J.}~\bibnamefont {Schaffner-Bielich}},\
  }\href {\doibase 10.1016/j.physletb.2014.02.028} {\bibfield  {journal}
  {\bibinfo  {journal} {Phys.Lett.}\ }\textbf {\bibinfo {volume} {B731}},\
  \bibinfo {pages} {154} (\bibinfo {year} {2014})}\BibitemShut {NoStop}%
\bibitem [{\citenamefont {Ferreira}\ \emph {et~al.}(2014)\citenamefont
  {Ferreira}, \citenamefont {Costa}, \citenamefont {Lourenço}, \citenamefont
  {Frederico},\ and\ \citenamefont {Providência}}]{Ferreira:2014kpa}%
  \BibitemOpen
  \bibfield  {author} {\bibinfo {author} {\bibfnamefont {M.}~\bibnamefont
  {Ferreira}}, \bibinfo {author} {\bibfnamefont {P.}~\bibnamefont {Costa}},
  \bibinfo {author} {\bibfnamefont {O.}~\bibnamefont {Lourenço}}, \bibinfo
  {author} {\bibfnamefont {T.}~\bibnamefont {Frederico}}, \ and\ \bibinfo
  {author} {\bibfnamefont {C.}~\bibnamefont {Providência}},\ }\href {\doibase
  10.1103/PhysRevD.89.116011} {\bibfield  {journal} {\bibinfo  {journal}
  {Phys.Rev.}\ }\textbf {\bibinfo {volume} {D89}},\ \bibinfo {pages} {116011}
  (\bibinfo {year} {2014})}\BibitemShut {NoStop}%
\bibitem [{\citenamefont {Farias}\ \emph {et~al.}(2014)\citenamefont {Farias},
  \citenamefont {Gomes}, \citenamefont {Krein},\ and\ \citenamefont
  {Pinto}}]{Farias:2014eca}%
  \BibitemOpen
  \bibfield  {author} {\bibinfo {author} {\bibfnamefont {R.}~\bibnamefont
  {Farias}}, \bibinfo {author} {\bibfnamefont {K.}~\bibnamefont {Gomes}},
  \bibinfo {author} {\bibfnamefont {G.}~\bibnamefont {Krein}}, \ and\ \bibinfo
  {author} {\bibfnamefont {M.}~\bibnamefont {Pinto}},\ }\href {\doibase
  10.1103/PhysRevC.90.025203} {\bibfield  {journal} {\bibinfo  {journal}
  {Phys.Rev.}\ }\textbf {\bibinfo {volume} {C90}},\ \bibinfo {pages} {025203}
  (\bibinfo {year} {2014})}\BibitemShut {NoStop}%
\bibitem [{\citenamefont {Ferrer}\ \emph
  {et~al.}(2014{\natexlab{a}})\citenamefont {Ferrer}, \citenamefont {de~la
  Incera},\ and\ \citenamefont {Wen}}]{Ferrer:2014qka}%
  \BibitemOpen
  \bibfield  {author} {\bibinfo {author} {\bibfnamefont {E.}~\bibnamefont
  {Ferrer}}, \bibinfo {author} {\bibfnamefont {V.}~\bibnamefont {de~la
  Incera}}, \ and\ \bibinfo {author} {\bibfnamefont {X.}~\bibnamefont {Wen}},\
  }\href@noop {} {\  (\bibinfo {year} {2014}{\natexlab{a}})},\ \Eprint
  {http://arxiv.org/abs/1407.3503} {arXiv:1407.3503 [nucl-th]} \BibitemShut
  {NoStop}%
\bibitem [{\citenamefont {Gies}\ and\ \citenamefont
  {Jaeckel}(2006)}]{Gies:2005as}%
  \BibitemOpen
  \bibfield  {author} {\bibinfo {author} {\bibfnamefont {H.}~\bibnamefont
  {Gies}}\ and\ \bibinfo {author} {\bibfnamefont {J.}~\bibnamefont {Jaeckel}},\
  }\href {\doibase 10.1140/epjc/s2006-02475-0} {\bibfield  {journal} {\bibinfo
  {journal} {Eur.Phys.J.}\ }\textbf {\bibinfo {volume} {C46}},\ \bibinfo
  {pages} {433} (\bibinfo {year} {2006})}\BibitemShut {NoStop}%
\bibitem [{\citenamefont {Mitter}\ \emph {et~al.}(2014)\citenamefont {Mitter},
  \citenamefont {Pawlowski},\ and\ \citenamefont
  {Strodthoff}}]{Mitter:2014wpa}%
  \BibitemOpen
  \bibfield  {author} {\bibinfo {author} {\bibfnamefont {M.}~\bibnamefont
  {Mitter}}, \bibinfo {author} {\bibfnamefont {J.~M.}\ \bibnamefont
  {Pawlowski}}, \ and\ \bibinfo {author} {\bibfnamefont {N.}~\bibnamefont
  {Strodthoff}},\ }\href@noop {} {\  (\bibinfo {year} {2014})},\ \Eprint
  {http://arxiv.org/abs/1411.7978} {arXiv:1411.7978 [hep-ph]} \BibitemShut
  {NoStop}%
\bibitem [{\citenamefont {Braun}\ and\ \citenamefont
  {Gies}(2007)}]{Braun:2005uj}%
  \BibitemOpen
  \bibfield  {author} {\bibinfo {author} {\bibfnamefont {J.}~\bibnamefont
  {Braun}}\ and\ \bibinfo {author} {\bibfnamefont {H.}~\bibnamefont {Gies}},\
  }\href {\doibase 10.1016/j.physletb.2006.11.059} {\bibfield  {journal}
  {\bibinfo  {journal} {Phys.Lett.}\ }\textbf {\bibinfo {volume} {B645}},\
  \bibinfo {pages} {53} (\bibinfo {year} {2007})}\BibitemShut {NoStop}%
\bibitem [{\citenamefont {Braun}\ and\ \citenamefont
  {Gies}(2006)}]{Braun:2006jd}%
  \BibitemOpen
  \bibfield  {author} {\bibinfo {author} {\bibfnamefont {J.}~\bibnamefont
  {Braun}}\ and\ \bibinfo {author} {\bibfnamefont {H.}~\bibnamefont {Gies}},\
  }\href {\doibase 10.1088/1126-6708/2006/06/024} {\bibfield  {journal}
  {\bibinfo  {journal} {JHEP}\ }\textbf {\bibinfo {volume} {0606}},\ \bibinfo
  {pages} {024} (\bibinfo {year} {2006})}\BibitemShut {NoStop}%
\bibitem [{\citenamefont {Scherer}\ and\ \citenamefont
  {Gies}(2012)}]{Scherer:2012nn}%
  \BibitemOpen
  \bibfield  {author} {\bibinfo {author} {\bibfnamefont {D.~D.}\ \bibnamefont
  {Scherer}}\ and\ \bibinfo {author} {\bibfnamefont {H.}~\bibnamefont {Gies}},\
  }\href {\doibase 10.1103/PhysRevB.85.195417} {\bibfield  {journal} {\bibinfo
  {journal} {Phys.Rev.}\ }\textbf {\bibinfo {volume} {B85}},\ \bibinfo {pages}
  {195417} (\bibinfo {year} {2012})}\BibitemShut {NoStop}%
\bibitem [{\citenamefont {Ferrer}\ \emph
  {et~al.}(2014{\natexlab{b}})\citenamefont {Ferrer}, \citenamefont {de~la
  Incera}, \citenamefont {Portillo},\ and\ \citenamefont
  {Quiroz}}]{Ferrer:2013noa}%
  \BibitemOpen
  \bibfield  {author} {\bibinfo {author} {\bibfnamefont {E.~J.}\ \bibnamefont
  {Ferrer}}, \bibinfo {author} {\bibfnamefont {V.}~\bibnamefont {de~la
  Incera}}, \bibinfo {author} {\bibfnamefont {I.}~\bibnamefont {Portillo}}, \
  and\ \bibinfo {author} {\bibfnamefont {M.}~\bibnamefont {Quiroz}},\ }\href
  {\doibase 10.1103/PhysRevD.89.085034} {\bibfield  {journal} {\bibinfo
  {journal} {Phys.Rev.}\ }\textbf {\bibinfo {volume} {D89}},\ \bibinfo {pages}
  {085034} (\bibinfo {year} {2014}{\natexlab{b}})}\BibitemShut {NoStop}%
\bibitem [{\citenamefont {Wetterich}(1993)}]{Wetterich:1992yh}%
  \BibitemOpen
  \bibfield  {author} {\bibinfo {author} {\bibfnamefont {C.}~\bibnamefont
  {Wetterich}},\ }\href {\doibase 10.1016/0370-2693(93)90726-X} {\bibfield
  {journal} {\bibinfo  {journal} {Phys.Lett.}\ }\textbf {\bibinfo {volume}
  {B301}},\ \bibinfo {pages} {90} (\bibinfo {year} {1993})}\BibitemShut
  {NoStop}%
\bibitem [{\citenamefont {Jungnickel}\ and\ \citenamefont
  {Wetterich}(1996)}]{Jungnickel:1995fp}%
  \BibitemOpen
  \bibfield  {author} {\bibinfo {author} {\bibfnamefont {D.}~\bibnamefont
  {Jungnickel}}\ and\ \bibinfo {author} {\bibfnamefont {C.}~\bibnamefont
  {Wetterich}},\ }\href {\doibase 10.1103/PhysRevD.53.5142} {\bibfield
  {journal} {\bibinfo  {journal} {Phys.Rev.}\ }\textbf {\bibinfo {volume}
  {D53}},\ \bibinfo {pages} {5142} (\bibinfo {year} {1996})}\BibitemShut
  {NoStop}%
\bibitem [{\citenamefont {Bethke}(2004)}]{Bethke:2004uy}%
  \BibitemOpen
  \bibfield  {author} {\bibinfo {author} {\bibfnamefont {S.}~\bibnamefont
  {Bethke}},\ }\href {\doibase 10.1016/j.nuclphysbps.2004.09.020} {\bibfield
  {journal} {\bibinfo  {journal} {Nucl.Phys.Proc.Suppl.}\ }\textbf {\bibinfo
  {volume} {135}},\ \bibinfo {pages} {345} (\bibinfo {year}
  {2004})}\BibitemShut {NoStop}%
\bibitem [{\citenamefont {Braun}(2012)}]{Braun:2011pp}%
  \BibitemOpen
  \bibfield  {author} {\bibinfo {author} {\bibfnamefont {J.}~\bibnamefont
  {Braun}},\ }\href {\doibase 10.1088/0954-3899/39/3/033001} {\bibfield
  {journal} {\bibinfo  {journal} {J.Phys.}\ }\textbf {\bibinfo {volume}
  {G39}},\ \bibinfo {pages} {033001} (\bibinfo {year} {2012})}\BibitemShut
  {NoStop}%
\bibitem [{\citenamefont {Braun}\ \emph {et~al.}(2011)\citenamefont {Braun},
  \citenamefont {Haas}, \citenamefont {Marhauser},\ and\ \citenamefont
  {Pawlowski}}]{Braun:2009gm}%
  \BibitemOpen
  \bibfield  {author} {\bibinfo {author} {\bibfnamefont {J.}~\bibnamefont
  {Braun}}, \bibinfo {author} {\bibfnamefont {L.~M.}\ \bibnamefont {Haas}},
  \bibinfo {author} {\bibfnamefont {F.}~\bibnamefont {Marhauser}}, \ and\
  \bibinfo {author} {\bibfnamefont {J.~M.}\ \bibnamefont {Pawlowski}},\ }\href
  {\doibase 10.1103/PhysRevLett.106.022002} {\bibfield  {journal} {\bibinfo
  {journal} {Phys.Rev.Lett.}\ }\textbf {\bibinfo {volume} {106}},\ \bibinfo
  {pages} {022002} (\bibinfo {year} {2011})}\BibitemShut {NoStop}%
\bibitem [{\citenamefont {Braun}\ \emph {et~al.}(2014)\citenamefont {Braun},
  \citenamefont {Fister}, \citenamefont {Pawlowski},\ and\ \citenamefont
  {Rennecke}}]{Braun:2014ata}%
  \BibitemOpen
  \bibfield  {author} {\bibinfo {author} {\bibfnamefont {J.}~\bibnamefont
  {Braun}}, \bibinfo {author} {\bibfnamefont {L.}~\bibnamefont {Fister}},
  \bibinfo {author} {\bibfnamefont {J.~M.}\ \bibnamefont {Pawlowski}}, \ and\
  \bibinfo {author} {\bibfnamefont {F.}~\bibnamefont {Rennecke}},\ }\href@noop
  {} {\  (\bibinfo {year} {2014})},\ \Eprint {http://arxiv.org/abs/1412.1045}
  {arXiv:1412.1045 [hep-ph]} \BibitemShut {NoStop}%
\bibitem [{\citenamefont {Gies}(2002)}]{Gies:2002af}%
  \BibitemOpen
  \bibfield  {author} {\bibinfo {author} {\bibfnamefont {H.}~\bibnamefont
  {Gies}},\ }\href {\doibase 10.1103/PhysRevD.66.025006} {\bibfield  {journal}
  {\bibinfo  {journal} {Phys.Rev.}\ }\textbf {\bibinfo {volume} {D66}},\
  \bibinfo {pages} {025006} (\bibinfo {year} {2002})}\BibitemShut {NoStop}%
\bibitem [{\citenamefont {Karbstein}\ \emph {et~al.}(2012)\citenamefont
  {Karbstein}, \citenamefont {Roessler}, \citenamefont {Dobrich},\ and\
  \citenamefont {Gies}}]{Karbstein:2011ja}%
  \BibitemOpen
  \bibfield  {author} {\bibinfo {author} {\bibfnamefont {F.}~\bibnamefont
  {Karbstein}}, \bibinfo {author} {\bibfnamefont {L.}~\bibnamefont {Roessler}},
  \bibinfo {author} {\bibfnamefont {B.}~\bibnamefont {Dobrich}}, \ and\
  \bibinfo {author} {\bibfnamefont {H.}~\bibnamefont {Gies}},\ }\href {\doibase
  10.1142/S2010194512007520} {\bibfield  {journal} {\bibinfo  {journal}
  {Int.J.Mod.Phys.Conf.Ser.}\ }\textbf {\bibinfo {volume} {14}},\ \bibinfo
  {pages} {403} (\bibinfo {year} {2012})}\BibitemShut {NoStop}%
\bibitem [{\citenamefont {Braun}\ \emph {et~al.}()\citenamefont {Braun},
  \citenamefont {Mian},\ and\ \citenamefont {Rechenberger}}]{BMR}%
  \BibitemOpen
  \bibfield  {author} {\bibinfo {author} {\bibfnamefont {J.}~\bibnamefont
  {Braun}}, \bibinfo {author} {\bibfnamefont {W.~A.}\ \bibnamefont {Mian}}, \
  and\ \bibinfo {author} {\bibfnamefont {S.}~\bibnamefont {Rechenberger}},\
  }\href@noop {} {\ }\Eprint {http://arxiv.org/abs/(in preparation)} {(in
  preparation)} \BibitemShut {NoStop}%
\bibitem [{\citenamefont {Aoki}\ \emph
  {et~al.}(2006{\natexlab{a}})\citenamefont {Aoki}, \citenamefont {Endrodi},
  \citenamefont {Fodor}, \citenamefont {Katz},\ and\ \citenamefont
  {Szabo}}]{Aoki:2006we}%
  \BibitemOpen
  \bibfield  {author} {\bibinfo {author} {\bibfnamefont {Y.}~\bibnamefont
  {Aoki}}, \bibinfo {author} {\bibfnamefont {G.}~\bibnamefont {Endrodi}},
  \bibinfo {author} {\bibfnamefont {Z.}~\bibnamefont {Fodor}}, \bibinfo
  {author} {\bibfnamefont {S.}~\bibnamefont {Katz}}, \ and\ \bibinfo {author}
  {\bibfnamefont {K.}~\bibnamefont {Szabo}},\ }\href {\doibase
  10.1038/nature05120} {\bibfield  {journal} {\bibinfo  {journal} {Nature}\
  }\textbf {\bibinfo {volume} {443}},\ \bibinfo {pages} {675} (\bibinfo {year}
  {2006}{\natexlab{a}})}\BibitemShut {NoStop}%
\bibitem [{\citenamefont {Aoki}\ \emph
  {et~al.}(2006{\natexlab{b}})\citenamefont {Aoki}, \citenamefont {Fodor},
  \citenamefont {Katz},\ and\ \citenamefont {Szabo}}]{Aoki:2006br}%
  \BibitemOpen
  \bibfield  {author} {\bibinfo {author} {\bibfnamefont {Y.}~\bibnamefont
  {Aoki}}, \bibinfo {author} {\bibfnamefont {Z.}~\bibnamefont {Fodor}},
  \bibinfo {author} {\bibfnamefont {S.}~\bibnamefont {Katz}}, \ and\ \bibinfo
  {author} {\bibfnamefont {K.}~\bibnamefont {Szabo}},\ }\href {\doibase
  10.1016/j.physletb.2006.10.021} {\bibfield  {journal} {\bibinfo  {journal}
  {Phys.Lett.}\ }\textbf {\bibinfo {volume} {B643}},\ \bibinfo {pages} {46}
  (\bibinfo {year} {2006}{\natexlab{b}})}\BibitemShut {NoStop}%
\bibitem [{\citenamefont {Cheng}\ \emph {et~al.}(2008)\citenamefont {Cheng},
  \citenamefont {Christ}, \citenamefont {Datta}, \citenamefont {van~der Heide},
  \citenamefont {Jung} \emph {et~al.}}]{Cheng:2007jq}%
  \BibitemOpen
  \bibfield  {author} {\bibinfo {author} {\bibfnamefont {M.}~\bibnamefont
  {Cheng}}, \bibinfo {author} {\bibfnamefont {N.}~\bibnamefont {Christ}},
  \bibinfo {author} {\bibfnamefont {S.}~\bibnamefont {Datta}}, \bibinfo
  {author} {\bibfnamefont {J.}~\bibnamefont {van~der Heide}}, \bibinfo {author}
  {\bibfnamefont {C.}~\bibnamefont {Jung}},  \emph {et~al.},\ }\href {\doibase
  10.1103/PhysRevD.77.014511} {\bibfield  {journal} {\bibinfo  {journal}
  {Phys.Rev.}\ }\textbf {\bibinfo {volume} {D77}},\ \bibinfo {pages} {014511}
  (\bibinfo {year} {2008})}\BibitemShut {NoStop}%
\bibitem [{\citenamefont {Mueller}\ and\ \citenamefont {Pawlowski}()}]{MP}%
  \BibitemOpen
  \bibfield  {author} {\bibinfo {author} {\bibfnamefont {N.}~\bibnamefont
  {Mueller}}\ and\ \bibinfo {author} {\bibfnamefont {J.~M.}\ \bibnamefont
  {Pawlowski}},\ }\href@noop {} {\ }\Eprint {http://arxiv.org/abs/(in
  preparation)} {(in preparation)} \BibitemShut {NoStop}%
\end{thebibliography}%

\end{document}